\documentstyle[aps,twocolumn,prl]{revtex}  

\begin{document}
\draft

\twocolumn[\hsize\textwidth\columnwidth\hsize\csname  
@twocolumnfalse\endcsname                             

\title{Andreev reflection at point contacts with heavy-fermion UBe$_{13}$ ?}

\author{K.~Gloos}
\address{Department of Physics, University of Jyv\"askyl\"a, 
  FIN-40351 Jyv\"askyl\"a, Finland}

\date{\today}
\maketitle



] 
\narrowtext 

In a recent Letter, W\"alti {\em et al.} \cite{Waelti00} have presented 
evidence for unconventional superconductivity in heavy-fermion UBe$_{13}$, 
using point-contact spectroscopy. 
They proposed that the huge zero-bias conductance peaks found for their 
contacts between UBe$_{13}$ and a Au tip are due to the existence of 
low-energy Andreev surface bound states indicating a  non-trivial energy-gap 
function. This interpretation implicitly assumes that the junctions are in
the ballistic limit, that means the electronic mean free path $l$ is
considerably larger than the contact radius $a$. 
As will be shown below, such a condition is unlikely to be fulfilled for 
contacts with UBe$_{13}$.
In the normal state  just above $T_c$, its electrical resistivity amounts 
to $\rho \approx 130\,\mu\Omega$cm, see for example 
Ref.~\cite{Rauchschwalbe87}.
This very large resistivity results in an extremely short electronic mean
free path $l \approx 1\,$nm, estimated using $\rho l = 
3\pi R_{\rm{K}} /2k_{\rm{F}}^{2}$ with $R_{\rm{K}} = h/e^2 = 
25.8\,{\rm{k}}\Omega$ and assuming a typical metallic Fermi wave number 
$k_{\rm{F}} \approx 10\,$nm$^{-1}$.

The basic properties of a metallic junction in the normal state are described
by Wexler's formula \cite{Wexler66}. Its approximate form splits up the 
contact resistance $R$ into a ballistic (also called Sharvin resistance) and 
a  resistive part (Maxwell resistance)
\begin{equation}
  R(T)  \approx  \frac {2R_{\rm{K}}} {(ak_{\rm{F}})^2}  
                                + \frac {\rho(T)}{{4a}}
\label{wexler}
\end{equation}
For simplicity, equal Fermi wave numbers with spherical Fermi surfaces 
on both sides of the junction are assumed, and the contribution of one of 
the electrodes (here the Au tip) to the resistive part has already been 
neglected. 
At large contacts with radii $a \gg l$, Maxwell's resistance $\rho(T)/4a$
dominates.
In the ballistic limit ($a \ll l$) the resistive part represents a 
small correction to the ballistic contact resistance, describing 
backscattering processes. 
At a resistance of
\begin{equation}
  R_{eq}  \approx  \frac {\left(\rho k_{\rm{F}}\right)^2} {16 R_{\rm{K}}} 
\label{resistance}
\end{equation}
both parts of the contact resistance have equal size. It requires $R \gg
R_{eq} \approx   410\,\Omega$ for a UBe$_{13}$ - Au junction to be in the
ballistic limit. This is indeed a very strong criterium  for spectroscopy
on these point contacts.

According to Wexler's formula Eq.~(\ref{wexler}), the resistive part vanishes
when the UBe$_{13}$ sample  becomes superconducting, leaving the ballistic 
part for possible Andreev reflection processes.
Therefore, any analysis in terms of Andreev reflection requires 
either the junction to be in the ballistic limit or to separate the different 
contributions to the superconducting signal.

Wexler's formula offers a  straightforward strategy to solve this
problem by comparing the temperature dependence of the specific 
resistivity in the  normal state with that of the contact resistance to 
derive the contact radius $a$. 
Akimenko {\em et al.} \cite{Akimenko82} first proposed and verified 
the principles of such an analysis on junctions between simple normal metals.
This method applied to junctions between UBe$_{13}$ and a normal 
metal (tungsten) showed the size of the superconducting anomalies to 
coincide with the resistive (Maxwell) part of the contact resistance 
over a wide range of contact radii $a$, indicating a negligible contribution 
of Andreev reflection \cite{Gloos96}.

The (typical) contact discussed in Ref.~\cite{Waelti00} has a normal-state 
$R \approx 2\,\Omega$.
Consequently, it is not in the ballistic limit: This resistance is mainly 
due to Maxwell's $\rho/4a$, with a contact radius $a \approx 160\,$nm,
while the ballistic part $2R_{\rm{K}} / (ak_{\rm{F}})^2$ is  quite small.
When the contact is cooled to below $T_c$, the electrical resistivity 
$\rho = 0$ and, thus, Maxwell's resistance vanishes. 
This leads to the observed large rise of the zero-bias conductance.
Approximating the ballistic resistance by the residual contact resistance 
$\sim 0.2\,\Omega$, the Fermi wave number $k_{\rm{F}} \approx 3\,$nm$^{-1}$,
a quite reasonable value. 

Of course, a more detailed investigation would also take into account 
that part of the contact region could stay normal due to the deformation 
of the UBe$_{13}$ crystal lattice caused by the Au tip, enhancing 
the residual contact resistance, or that there could be contributions from 
Andreev reflection. 
However, the latter processes are difficult to identify because of the 
large resistive (Maxwell) component of the superconducting signal.


\end{document}